\documentclass[12pt]{article}
\usepackage[esperanto]{babel}
\usepackage{epsfig} 
\usepackage{icomma}

\newcommand{\dd}{\mathrm{d}}

\newcommand{\bea}{\vspace{-1ex}\begin{eqnarray}}
\newcommand{\eea}{\end{eqnarray}}

\title{{\bf Relativeca Dopplera efiko \\ \^ce unuforme akcelata movo -- III}}

\author{F.M. Paiva \\ 
{\small Departamento de F\'\i sica, U.E. Humait\'a II, Col\'egio Pedro II} \\
{\small Rua Humait\'a 80, 22261-040  Rio de Janeiro-RJ, Brasil; fmpaiva@cbpf.br} 
\vspace{.7ex} \\
A.F.F. Teixeira \\
{\small Centro Brasileiro de Pesquisas F\'\i sicas} \\
{\small 22290-180 Rio de Janeiro-RJ, Brasil; teixeira@cbpf.br}} 

\begin{document}

\selectlanguage{esperanto}
\maketitle 
\thispagestyle{empty}

\begin{abstract}  
^Ce special-relativeco ni detale priskribas lum-Doppleran efikon inter rest\-anta lum-fonto kaj observanto kun rektlinia movado kaj konstanta propra akcelo. 

In the context of special relativity, we describe with detail the Doppler effect between a light source at rest and an observer in linear motion and constant proper acceleration. To have an English version of this article, ask the authors.  
\end{abstract}

\section{\label{e}Enkonduko}

En anta^uaj artikoloj~\cite{PaivaTeixeira2006,PaivaTeixeira2007a, PaivaTeixeira2007b, PaivaTeixeira2008a} ni pridiskutis Doppleran efikon de lum-signalo inter fonto kaj observanto. Tie a^u amba^u movi^gas a^u unu movi^gas dum la alia restas. Ofte estas komenca fazo, kiam movi^ganta observanto ricevas signalon eligitan kiam fonto ankora^u ne movi^gas. La Doppleran faktoron de tia fazo ni kalkulis en~\cite{PaivaTeixeira2008a}, kaj nun ni plu esploras ties grafika^jojn.

\begin{figure}                                                
\centerline{\epsfig{file=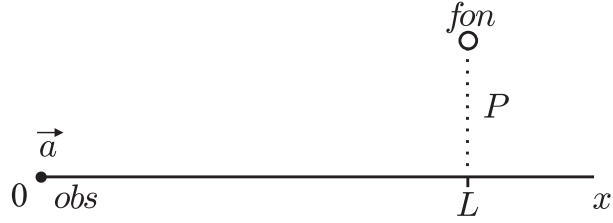,width=8cm}} 
\caption{Komencaj lokoj de observanto kaj fonto.}
\label{fig.komencaj} 
\end{figure}

Tia sistemo fari^gas kun observanto ko\-men\-ce en loko $[x, y] = [0, 0]$, kaj lum-fonto en $[L, P]$, amba^u restantaj. ^Ci tie $L$ estas a^u pozitiva, a^u nula a^u malpozitiva, kaj tute-^generale  $P$ estas a^u pozitiva a^u nula, la^u figuro~\ref{fig.komencaj}. Ekde momento $t=0$, la observanto movi^gas kun konstanta propra akcelo $a$, pozitive de akso $x$, dum la fonto plu restas. Rimarku, ke se $L>0$, la observanto komence proksimi^gas al la fonto, kaj se $L<0$ ^gi ^ciam fori^gas de la fonto.  La^u~\cite[ekvacio~(23)]{PaivaTeixeira2008a}, por $t>0$ la Dopplera faktoro estas

\bea                                                       \label{ek.Dop}
D(\tau)=
\cosh(a\tau/c) + \frac
      {[1 + l - \cosh(a\tau/c)]\,\sinh(a\tau/c)}
{\sqrt{[1 + l - \cosh(a\tau/c)]^2 + p^2}}\ , 
\hspace{1mm} l:=aL/c^2, \hspace{1mm} p:=aP/c^2\ ,
\eea 
kie $\tau$ estas la propra tempo de observanto, elektante $\tau=0$ kiam $t=0$.

Sekve ni priskribas tri gravajn aferojn rilatajn al nia studo: tempan dilaton, propran akcelon, kaj Doppleran efikon.

\subsubsection{Tempa dilato}

Esti^gu inercia referenca sistemo $S$, t.e., unu normohorlo^go fiksa en ^ciu spaca punkto, ^ciuj sinkronaj. La tempo $t$ mezurata per tiu horlo^garo nomi^gas tempa koordinato. Plu esti^gu movi^ganta normohorlo^go $\cal H$, kaj $\tau$ ^gia propra tempo. 

Supozu, ke kiam $\cal H$ montras tempon $\tau_1$, ^gi trapasas horlo^gon de $S$ kiu montras tempon $t_1$. Poste, kiam $\cal H$ montras tempon $\tau_2$, ^gi trapasas alian horlo^gon de $S$, kiu montras tempon $t_2$. La^u special-relativeco la koordinata intertempo $t_2-t_1$ mezurata per la horlo^garo de $S$ estas pli granda ol la propra intertempo $\tau_2-\tau_1$ mezurata per horlo^go $\cal H$. Se intertempoj $t_2-t_1=:\dd t$ kaj $\tau_2-\tau_1=:\dd \tau$ estas infinitezimaj, do  special-relativeco diras, ke $\dd t = \gamma\dd \tau$, kie $\gamma:=1/\sqrt{1-v^2/c^2}$, estante $v$ la rapido de $\cal H$. ^Car  ^ciam $\gamma\geq1$, do ^ciam $\dd t\geq\dd \tau$, kaj tiu ebla malsameco nomi^gas tempa dilato.

\subsubsection{Propra akcelo}

Special-relativeco ne permesas, ke objekto havas konstantan\, Newtonan akcelon $a_N:=\dd^2x/\dd t^2$ dum tro longa tempo. Fakte, tia objekto atingus rapidon $v:=\dd x/\dd t$ pli granda ol vakuo-lumo-rapido $c$, kio ne estas ebla en tiu teorio. 

Plej ta^ugas al special-relativeco propra akcelo $a:=\gamma^3a_N$. Memoru, ke ju pli granda la rapido des pli granda la faktoro de tempa dilato $\gamma$, kaj kiam la rapido proksimi^gas al $c$, tiam $\gamma$ emas al $\infty$. Rezultas en ^ci tiu studo, ke konstanta propra akcelo $a$ respondas al malkreskanta Newtona akcelo $a_N$ kaj, ke la rapido de objekto kun konstanta propra akcelo dum tre longa tempo apena^u emas al vakuo-lumo-rapido~$c$. 

Fizike, oni povas difini propran akcelon $a$ de objekto kiel la Newtona akcelo mezurata en inercia referenca sistemo kun momente la sama rapido de objeto. Studoj de movado kun konstanta $a$ estas oftaj, kiel ^ce M\o ller \cite[pa^go 72]{Moller}, Rindler \cite[pa^go 49]{Rindler}, Dwayne Hamilton \cite{Hamilton}, Landau kaj Lifshitz \cite[pa^go 22]{LandauLifshitz}, Cochran \cite{Cochran}, kaj ni mem \cite{PaivaTeixeira2006, PaivaTeixeira2007a, PaivaTeixeira2007b, PaivaTeixeira2008a}.

\subsubsection{Dopplera efiko}

En propra momento $\tau_f$ fonto eligas lum-signalon kun frekvenco $\nu_f$, la^u siaj propraj mezuriloj. En propra momento $\tau$ observanto ricevas tiun lum-signalon kun frekvenco $\nu$, la^u siaj propraj mezuriloj. Plej ofte $\nu\neq\nu_f$; ^ci tiu ebla ^san^go de frekvenco nomi^gas Dopplera efiko, kaj la kvociento $D(\tau):=\nu/\nu_f$ nomi^gas Dopplera faktoro. Se $\nu>\nu_f$, la fenomeno nomi^gas al-violo ($D>1$); kontra^ue nomi^gas al-ru^go ($D<1$). 

La^u~\cite{PaivaTeixeira2007a, PaivaTeixeira2007b, PaivaTeixeira2008a}, la Doppleran faktoron oni povas kalkuli konsiderante du infinitezime sinsekvajn lum-signalojn. Esti^gu $\dd\tau_f$ la intertempo de eligo de tiuj du signaloj mezurita per la fonto, kaj esti^gu $\dd\tau$ la intertempo de ricevo de tiuj du signaloj mezurita per la observanto. Ekvivalente, esti^gu $\dd t_f$ kaj $\dd t$ la respondaj intertempoj, mezuritaj per iu inercia referenca sistemo. La Dopplera faktoro je la momento $\tau$ de ricevo de signalo estas

\bea
\label{ek.^generala}
D(\tau)=
\frac{\dd\tau_f}{\dd\tau} = 
\frac{\gamma(t)}{\gamma_f(t_f)}
\frac{\dd t_f}{\dd t}\ .
\eea

En (\ref{ek.^generala}), la faktoro de tempa dilato $\gamma(t)$ pro movado de observanto kontribuas por alviolo, kaj, en la nomanto, la faktoro de tempa dilato $\gamma_f(t_f)$ pro movado de fonto kontribuas por alru^go. La fori^g-proksimi^ga faktoro $\dd t_f/\dd t$ povas kontribui amba^umaniere: dum proksimi^go ^gi estas pli granda ol $1$, do kontribuas por alviolo, kaj dum fori^go ^gi estas pli eta ol $1$, do kontribuas por alru^go. La Dopplera faktoro~(\ref{ek.^generala}) estas rezulto de tiuj tri kontribuoj.

En nuna studo la fonto restas, do $\gamma_f(t_f)=1$, kaj do $D(\tau)=\gamma(t)\,\dd t_f/\dd t$ havas nur du faktorojn. Tio simpligas la diskuton pri ekvacio~(\ref{ek.^generala}) al la jenaj ^generala^joj.  1:~^ce proksimi^go de observanto al fonto, la Dopplera efiko estas alviolo pro amba^u faktoroj. 2:~^ce preterpaso, t.e., pozicio $x=L$ de  movi^ganta observanto en figuro~\ref{fig.komencaj} estante $P\neq0$, faktoro $\dd t_f/\dd t=1$ ne kontribuas, do $\gamma(t)$ faktoro estras farante alviolon. 3:~^ce fori^go, a^u la alru^ga faktoro $\dd t_f/\dd t$ a^u la alviola faktoro $\gamma(t)$ estras, farante plurajn interesajn eblecojn. 

Sekcio~\ref{sek.Grafo} montras nian plej gravan rezulton, la studon de grafika^joj de Dopplera faktoro.  Sekcio~\ref{sek.detalo} montras iujn detalojn plu, kaj Sekcio~\ref{sek.konkludo} konkludas.

\section{\label{sek.Grafo}Grafika^joj} 

\begin{figure}                                               
\centerline{\epsfig{file=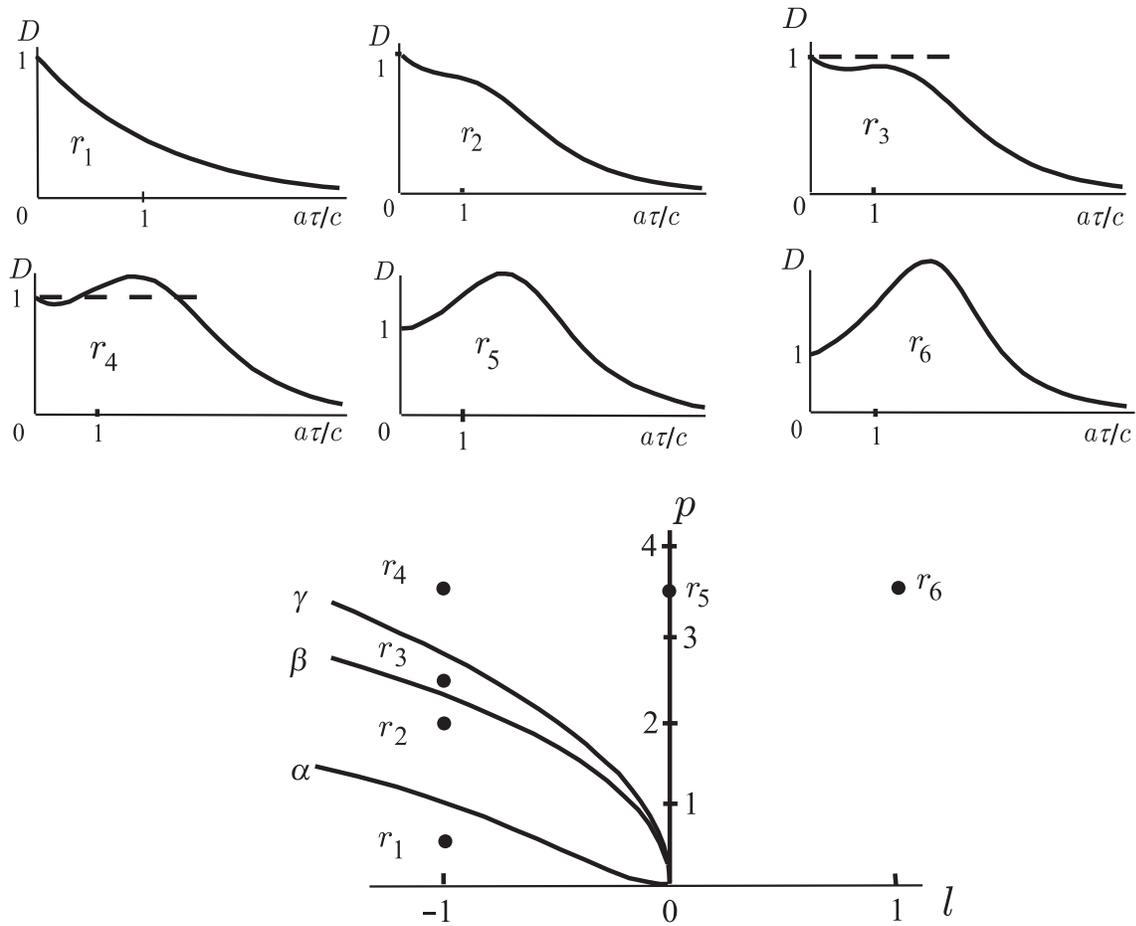,width=15cm}} 
\caption{Super estas 6 ecoj de grafika^jo de Dopplera\, faktoro\, $D$ kontra^u $a\tau/c$, respondaj al ^ciu regiono de la malsupera diagramo. Tiu diagramo montras 6 regionojn, de $r_1$ ^gis $r_6$, la^u valoroj de $l=aL/c^2$ kaj $p=aP/c^2$, kaj montras kurbojn \{$\alpha,\beta,\gamma$\}, kiuj limigas tiujn regionojn. Regiono $r_5$ estas akso $l=0$. Nigraj buloj ($\bullet$) montras la valorojn de $l$ kaj $p$ rilatajn al ^ciu grafika^jo.}
\label{fig.plejgrava} 
\end{figure}

Nun ni studas grafika^jojn de Dopplera faktoro el ekvacio~(\ref{ek.Dop}), montratajn en figuro~\ref{fig.plejgrava}. \^Car en komenca momento $\tau=0$ la observanto restas, tiam Dopplera efiko ne estas, $D(0)=1$. Laste, kiam $\tau\rightarrow\infty$, okazas fori\^go, kaj la interspaco de fonto al observanto estas multe pligranda ol $P$. Do la movado estas preska\u u kolinia, kaj do la Dopplera efiko emas al plej forta alru\^go, t.e., la Dopplera faktoro estas $D(\infty)\rightarrow0$.

La signaloj de $D\,'(0)$ kaj $l$ estas la sama. Tio estas facile komprenebla. Ja, se $l>0$ la movado estas komence proksimi^go, do la Dopplera efiko estas alviolo, kaj se $l<0$ la movado estas komence fori^go kun malgranda rapido, do la Dopplera efiko estas alru^go. Anka^u rimarku, ke $D\,''(0)>0$, do la unua deriva^jo $D\,'(\tau)$ plii^gas je $\tau=0$.

Nun ni analizas ^ciu grafika^jo en figuro~\ref{fig.plejgrava}. Estas ses ecoj de grafika^jo, pendante de valoroj de $l$ kaj $p$. En tiu figuro, malsupera diagramo $l\times p$ limigas ses regionojn, de $r_1$ ^gis $r_6$, per kurboj \{$\alpha,\beta,\gamma$\}. ^Ciu regiono respondas al iu eco de grafika^jo.

Grafika^joj de regiono $r_1$ estas la plej simplaj: ^ciam $D(\tau)\leq1$, $D\,'(\tau)<0$ kaj $D''(\tau)>0$, do ili similas al $\exp(-a\tau/c)$. Vere, ili estas ^guste $D(\tau)=\exp(-a\tau/c)$ se $p=0$ kaj $l<0$. 

Kurbo $\alpha$ estas limo inter regionoj $r_1$ kaj $r_2$, t.e., inter regiono de $D\,''(\tau)$ ^ciam pozitiva kaj regiono kie ^gi povas esti  malpozitiva. Tiu kurbo malkovri^gas per forigo de $\tau$ el  $D\,''(l,p,\tau)=0$ kaj $D\,'''(l,p,\tau)=0$. Oni montras, ke $\alpha$ havas horizontalan asimptoton, $p^2 = \frac{3}{2}(11\sqrt{33}-59)$, t.e., $p\approx 2,51$.

Grafika^joj de $r_2$ malsami^gas al tiuj de $r_1$ pro ^gibo. Tamen tiu ^gibo ne havas maksimumon, ^car ^ciam $D\,'(\tau)\neq0$. Tiu ^gibo okazas pro pliforta alviola kontribuo de $\gamma$ faktoro. Vere, ^car $p$ estas pli granda kompare kun $r_1$, komence la fori^go estas malpli kolinia ol en $r_1$, do alru^ga kontribuo de fori^ga faktoro estas malpli forta. Tamen, tiu alru^ga kontribuo pliforti^gas kiam la fori^go estas preska^u kolinia.

Kurbo $\beta$ estas limo inter regionoj $r_2$ kaj $r_3$, t.e., inter regiono de $D\,'(\tau)$ ^ciam malpozitiva kaj regiono kie ^gi povas esti pozitiva. Tiu kurbo malkovri^gas per forigo de $\tau$ el ekvacioj   $D\,'(l,p,\tau)=0$ kaj $D''(l,p,\tau)$ $=0$. Jen $\beta$:

\bea                                                     \label{ek.beta}
l = \frac{1-5\,p^2/27}{\sqrt{1-p^2/27}}-1\ .
\eea
Oni klare rimarkas el (\ref{ek.beta}), ke $\beta$ havas horizontalan asimptoton $p=3\sqrt{3}$.

Grafika^joj de $r_3$ malsami^gas al tiuj de $r_2$ ^car la ^gibo havas maksimumon. Tamen tiu maksimumo estas pli eta ol $1$, do alviolo ne okazas. Ju pli granda $p$ kompare kun $|l|$ des malpli kolinia la komenca fori^go. Do la alru^ga kontribuo de fori^ga faktoro malforti^gas komence kaj la ^gibo plii^gas.

Kurbo $\gamma$ estas limo inter regionoj $r_3$ kaj $r_4$, t.e., inter regiono de $D(\tau)$ pli eta ol $1$ kaj regiono kie ^gi povas esti pli granda ol $1$. Tiu kurbo malkovri^gas per forigo de $\tau$ el ekvacioj $D(l,p,\tau)=1$ kaj $D\,'(l,p,\tau)=0$; ^gi estas la parabolo $p = 2\sqrt{-2\,l\,}$.  

Grafika^joj de $r_4$ malsami^gas al tiuj de $r_3$ ^car iliaj maksimumoj estas pli granda ol $1$, t.e., poste komenca alru^go estas alviolo. Tio okazas, ^car la fori^ga movado komence estas sufi^ce ne kolinia, tiel kiel, post iom da tempo, alviola kontribuo de $\gamma$ faktoro superas la alru^ga kontribuo de fori^ga faktoro. Tio estas senpere relativeca fenomeno, ^car Newtone nur alru^go okazas ^ce fori^ga movado.

Grafika^joj de $r_5$ estas en limo inter $r_4$ kaj $r_6$. Ili respondas al $l=0$ do havas $D\,'(0)=0$. Tio okazas ^car komence ne estas proksimi^go nek fori^go. Tuj poste, alviola kontribuo de $\gamma$ faktoro estras super alru^ga kontribuo de fori^ga faktoro. Laste, la movado fari^gas preska^u kolinia fori^go, do alru^go de fori^ga faktoro estras.   

Grafika^joj de $r_6$ respondas al $l>0$. 
Komence estas proksimi^go de observanto al fonto, do proksimi^ga faktoro $\dd t_f/\dd t$ fortigas la alviolan kontribuon de $\gamma$. Laste, la movado fari^gas preska^u kolinia fori^go, kaj la alru^ga kontribuo de fori^ga faktoro estras.

\section{\label{sek.detalo}Detalo}

\begin{figure}                                                
\centerline{\epsfig{file=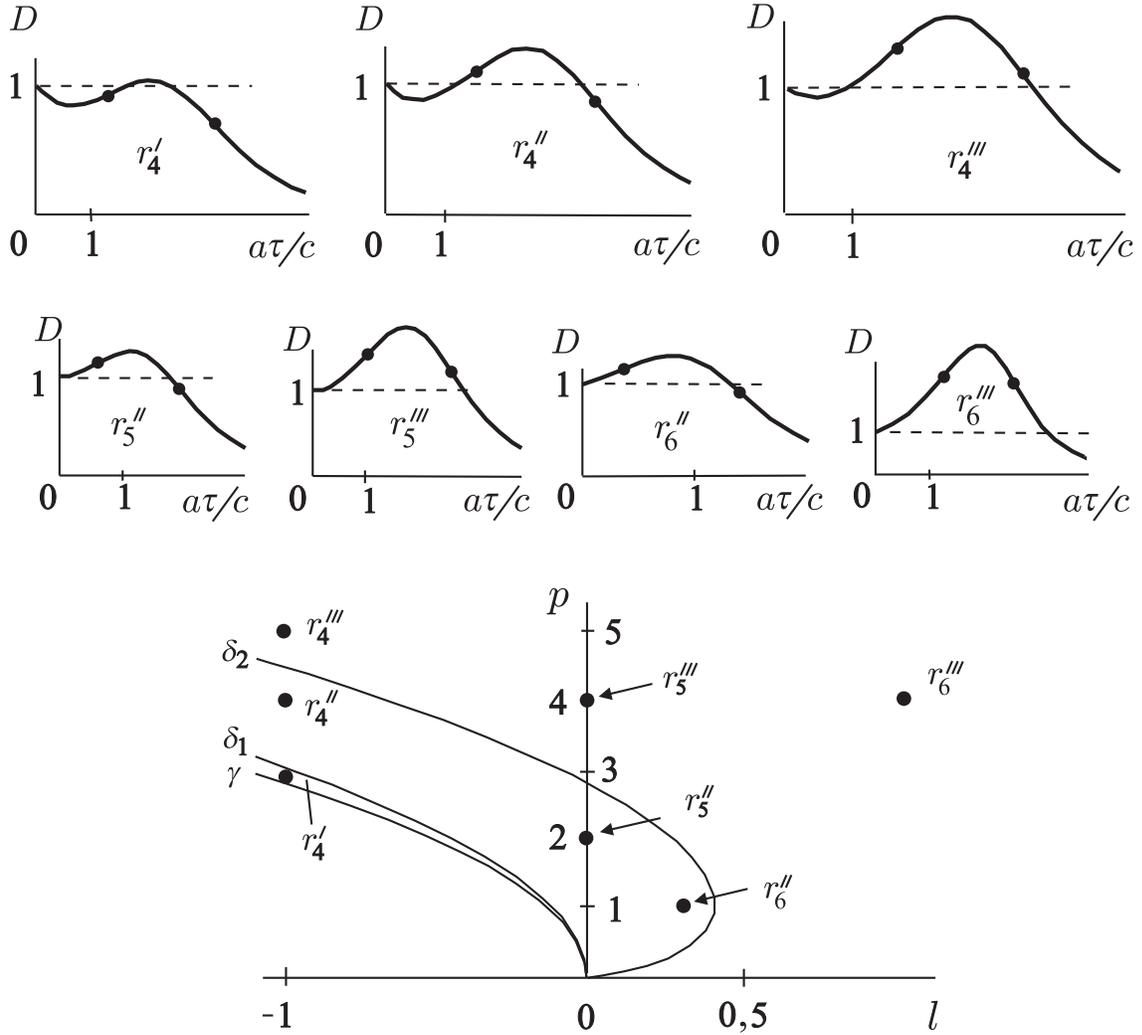,width=15cm}} 
\caption{Plu detaloj pri regionoj $r_4$, $r_5$ kaj $r_6$ el figuro~\ref{fig.plejgrava}. Super estas sep ecoj de grafika^jo de Dopplera faktoro $D$ kontra^u $a\tau/c$, responde al ^ciu subregiono de la malsupera diagramo. Tiu diagramo montras sep subregionojn $\{r'_4$, $r''_4$, $r'''_4$, $r''_5$, $r'''_5$, $r''_6$,  $r'''_6\}$ la^u valoroj de $l=aL/c^2$ kaj $p=aP/c^2$, kaj montras kurbojn $\gamma$, $\delta_1$ kaj $\delta_2$, kiuj limigas tiuj subregionoj. Subregionoj $r''_5$ (havante $p<2\sqrt{2}$) kaj $r'''_5$ (kies $p>2\sqrt{2}$) estas sur akso $l=0$. Nigraj buloj ($\bullet$) en la diagramo montras la valorojn de $l$ kaj $p$ rilatajn al ^ciu grafika^jo, kaj nigraj buletoj ({\scriptsize\textbullet}) en la grafika^joj montras transfleksejojn.} 
\label{fig.detaloj} 
\end{figure}

Eblas pli detala analizo de regionoj $r_4$, $r_5$ kaj $r_6$ el figuro~\ref{fig.plejgrava}. Fakte, figuro~\ref{fig.detaloj} montras subregionojn $\{r'_4$, $r''_4$, $r'''_4\}$, $\{r''_5$, $r'''_5\}$, $\{r''_6$, $r'''_6\}$, la^u transfleksejoj en grafika^joj estas super a^u sub $D=1$.

Kurboj $\delta_1$ kaj $\delta_2$ limigas tiujn subregionoj kaj rilatas al la unua kaj dua transfleksejo, respektive. Amba^u kurboj malkovri^gas per forigo de $\tau$ el ekvacioj $D(l,p,\tau)=1$ kaj $D''(l,p,\tau)=0$. Uzante la parametron $m\in[1,\infty]$, jen ili:
\bea                                                 \label{ek.delta}
l = \frac{m-1}          {m+1}\left(1-2m\mp\sqrt{6+3m^2}\right),
\hspace{2em} 
p = \frac{\sqrt{2(m-1)}}{m+1}\left(  3m\pm\sqrt{6+3m^2}\right)\ ,
\eea
kie la superaj kaj subaj signaloj respondas al $\delta_1$ kaj  $\delta_2$, respektive. Oni montras, ke amba^u kurboj kuni^gas kun parabolo \mbox{$p=2\sqrt{-3\,l\,}$} ^ce $m\rightarrow\infty$.

\section{\label{sek.konkludo}Konkludo}

Certe oni povas fari pli detalan matematikan analizon de grafika^joj de Dopplera faktoro kaj pridiskuti interpreton el ili. Ekzemple, akso $p=0$ el regiono $r_6$ respondas al trapaso de fonto per observanto, farante malkontinuecon en Dopplera faktoro. Tiun aferon studas~\cite{PaivaTeixeira2007a} kaj la plej grava rezulto estas en ties figuro~{\bf 5.b}, plej specife en la kvadranto kun $\tau>0$. La malkontinueco okazas ^car je la momento de trapaso, alviolo pro kolinia proksimi^go subite fari^gas alru^go pro kolinia fori^go. 

Interesas rimarki, ke ^ci tiu studo ta^ugas anka^u por komenca fazo de Dopplera efiko inter du akcelataj korpoj, t.e., kiam movi^ganta observanto ricevas signalon el fonto, kiu ankora^u ne movi^gas. Tiun problemon pridiskutas anta^ua artikolo~\cite{PaivaTeixeira2008a}. En estonta artikolo ni uzas rezultojn de ^ci~tiu nuna artikolo por pli detaligi rezultojn de~\cite{PaivaTeixeira2008a}.

\selectlanguage{esperanto}

\end{document}